\documentclass[10pt,twocolumn,letterpaper]{article}
\pdfoutput=1

\usepackage[hyphens]{url}
\usepackage[left=1in,right=1in,top=1in,bottom=1in,
            columnsep=0.25in,nohead
            ]{geometry}

\makeatletter

\def\ftype@copyrightbox{8}
\def\@copyrightspace{
\@float{copyrightbox}[b]
\begin{center}
\setlength{\unitlength}{1pc}
\begin{picture}(20,7.0) 
\put(0,3){\parbox{\columnwidth}{\footnotesize

}
}
\end{picture}
\end{center}
\end@float}

\def\maketitle{\par
 \begingroup
     \twocolumn[\@maketitle]
\@thanks
 \endgroup
 \setcounter{footnote}{0}
 \let\maketitle\relax
 \let\@maketitle\relax
 \gdef\@thanks{}\gdef\@author{}\gdef\@title{}\gdef\@subtitle{}\let\thanks\relax
}

\makeatother

\usepackage{times} 
\usepackage{mathptmx}
\usepackage{amsmath,amsfonts,amssymb,graphicx,color,times,amsthm,url}
\usepackage[noend]{algorithmic}
\usepackage{lastpage}
\usepackage{subfig}

\theoremstyle{definition}

\newcommand{\esrc}{\mathit{src}}
\newcommand{\edst}{\mathit{dst}}
\newcommand{\eime}{\mathit{time}}
\newcommand{\infilter}{\mathbin{@}}
\newcommand{\outfilter}{\mathbin{\backslash\!\!\!\!\!\!\!{@}}}

\newcommand{\DClos}{\downarrow}

\newcommand{\msgon}[1]{\Pi_{#1}}

\newcommand{\Ine}{\mathit{In}_e}
\newcommand{\Oute}{\mathit{Out}_e}
\newcommand{\processedmess}{\bar{M}}
\newcommand{\processednot}{\bar{N}}
\newcommand{\discardedmess}{\bar{D}}
\newcommand{\CMetadata}{\Xi}

{\end{list}}

{\end{list}}

\newcounter{@inst}
\newcounter{@auth}

\newdimen\instindent
\newbox\authrun
\newtoks\authorrunning
\newtoks\tocauthor
\newbox\titrun
\newtoks\titlerunning
\newtoks\toctitle

\def\clearheadinfo{\gdef\@author{No Author Given}%
                   \gdef\@title{No Title Given}%
                   \gdef\@subtitle{}%
                   \gdef\@institute{No Institute Given}%
                   \gdef\@thanks{}%
                   \global\titlerunning={}\global\authorrunning={}%
                   \global\toctitle={}\global\tocauthor={}}

\def\institute#1{\gdef\@institute{#1}}

\def\institutename{\par
 \begingroup
 \parskip=\z@
 \parindent=\z@
 \setcounter{@inst}{1}%
 \def\and{\par\stepcounter{@inst}%
 \noindent$^{\the@inst}$\enspace\ignorespaces}%
 \setbox0=\vbox{\def\thanks##1{}\@institute}%
 \ifnum\c@@inst=1\relax
   \gdef\fnnstart{0}%
 \else
   \xdef\fnnstart{\c@@inst}%
   \setcounter{@inst}{1}%
   \noindent$^{\the@inst}$\enspace
 \fi
 \ignorespaces
 \@institute\par
 \endgroup}

\def\@fnsymbol#1{\ensuremath{\ifcase#1\or\star\or{\star\star}\or
   {\star\star\star}\or \dagger\or \ddagger\or
   \mathchar "278\or \mathchar "27B\or \|\or **\or \dagger\dagger
   \or \ddagger\ddagger \else\@ctrerr\fi}}

\begin{document}

\title{\bf Falkirk Wheel: Rollback Recovery for Dataflow Systems}
\author{Michael Isard and Mart\'in Abadi: DRAFT paper work in progress}
\date{}
\maketitle 
\thispagestyle{empty}

\begin{abstract}
  
  We present a new model for rollback recovery in distributed dataflow
  systems. We explain existing rollback schemes by assigning a logical
  time to each event such as a message delivery. If some processors
  fail during an execution, the system rolls back by selecting a set
  of logical times for each processor. The effect of events at times
  within the set is retained or restored from saved state, while the
  effect of other events is undone and re-executed.  We show that, by
  adopting different logical time ``domains'' at different processors,
  an application can adopt appropriate checkpointing schemes for
  different parts of its computation. We illustrate with an example of
  an application that combines batch processing with low-latency
  streaming updates. We show rules, and an algorithm, to determine a
  globally consistent state for rollback in a system that uses
  multiple logical time domains. We also introduce \emph{selective
    rollback} at a processor, which can selectively preserve the
  effect of events at some logical times and not others, independent
  of the original order of execution of those events.  Selective
  rollback permits new checkpointing policies that are particularly
  well suited to iterative streaming algorithms. We report on an
  implementation of our new framework in the context of the Naiad
  system.

\end{abstract}

\section{Introduction}
\label{sec:intro}

This paper is about fault tolerance in distributed dataflow systems.
Specifically, we investigate the information that must be tracked and
persisted in order to restart a system in a consistent state after the
failure of one or more processes.  We assume other requirements, such
as detecting failures and reliably persisting state, are adequately
covered by existing techniques. We describe a general mechanism and an
implementation of it in the context of the
Naiad~\cite{murray:naiad:sosp13} system. We also suggest how the ideas
may be applied to other distributed systems. The mechanism is named
after the Falkirk Wheel~\cite{FalkirkWheel}, a prior engineering
solution for high-throughput streaming rollback.

\begin{figure}[tbp]%
  \begin{center}%
    \leavevmode
    \includegraphics[width=8cm]{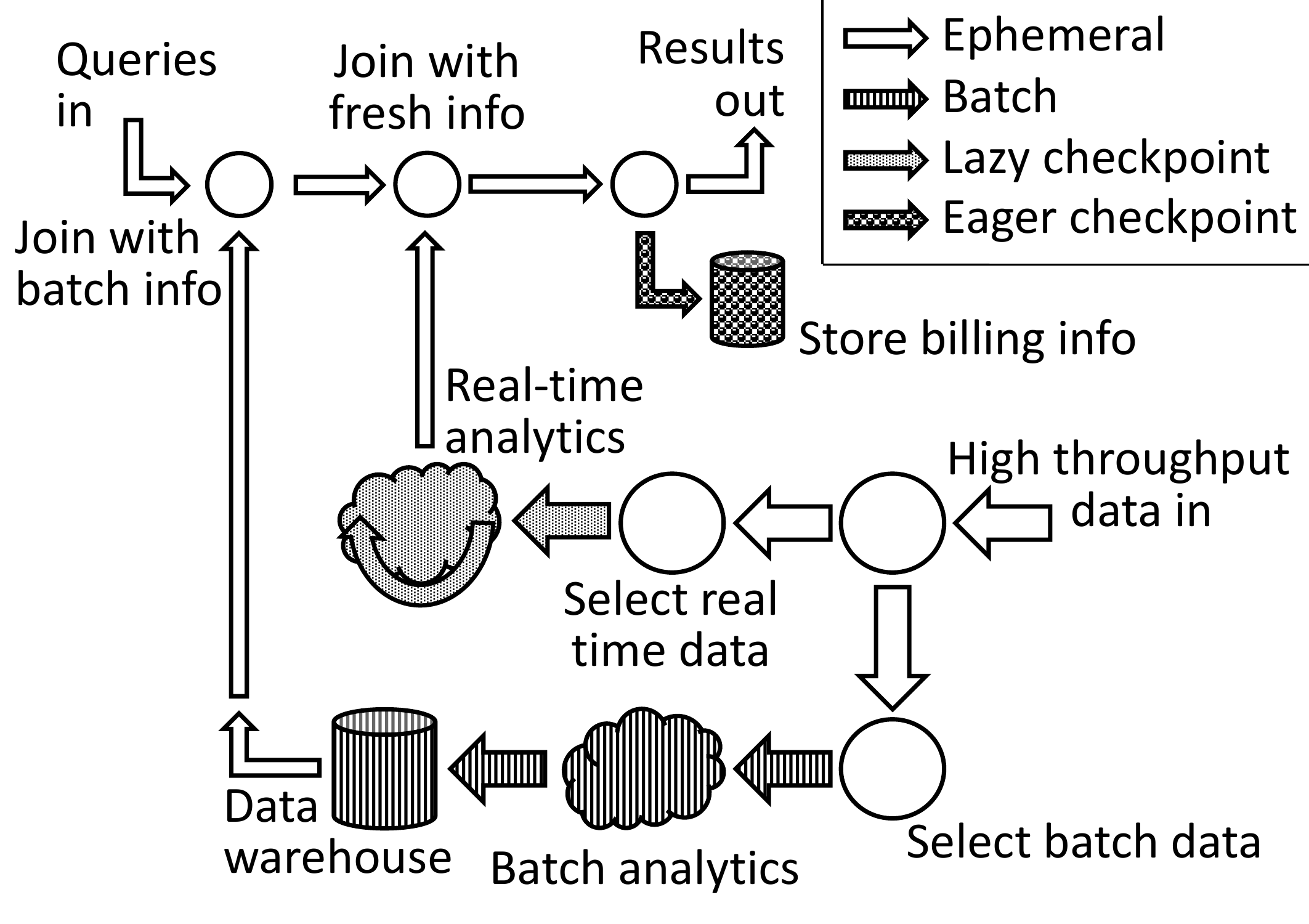}
  \end{center}\vskip -3ex%
  \caption{A complex streaming application. Different parts of the
    computation have different availability, throughput and latency
    requirements, and thus merit different fault-tolerance
    policies.}\vskip -2ex
  \label{fig:scenario}
\end{figure}
Most fault-tolerant distributed systems adopt a fixed policy for
checkpointing and logging. As a result, all applications running on
these systems must operate with the same set of performance tradeoffs.
Streaming applications often require high availability, i.e.\ the
system must resume output soon after the detection of a failure.
Systems designed for these applications must be able to restore
quickly to a recent consistent state on failure, meaning they must
frequently update persistent state. Other applications may be more
sensitive to throughput or latency, which are hard to maintain while
eagerly writing to stable storage.  These conflicting application
requirements are a major motivation for the development of multiple
systems such as Spark~\cite{zaharia:spark:nsdi12}, Storm~\cite{storm},
S4~\cite{S4} and Millwheel~\cite{akidau:millwheel:vldb13}. We argue
that such systems would be more useful if they could mix policies, and
thus performance tradeoffs, within a single application.

Consider the application in Figure~\ref{fig:scenario}.  User queries
arrive at the top left and are joined with two sets of data: the
output of a periodic batch computation; then the output of a
continuously-updated iterative computation. Statistics about the query
response are then stored in a database and the response is delivered
back to the user. Concurrently the application receives a
high-throughput stream of data records. Some fields of these records
are directed to the batch computation, which is re-run periodically.
Other fields are used as inputs to the iterative computation which
updates in real time.

The application adopts four separate fault-tolerance regimes for
different regions of the computation, indicated by the different
shading used for different parts of the dataflow illustration. The
first we call ``ephemeral'' which means that the records flowing
through this part of the graph are never saved to stable storage, and
none of the dataflow vertices they pass through store mutable state.
Clients that introduce ephemeral records (users sending queries or the
external service supplying high-throughput data) do not receive an
acknowledgement until the records have flowed through the entire
ephemeral subgraph, so fault tolerance for these records is attained
by requiring clients to retry on failure. Data reductions are
performed on the high-throughput input records before they leave the
ephemeral regime.  The second regime is ``batch.''  In this part of
the graph there is a high-throughput data-intensive computation that
is run periodically and can tolerate re-execution that introduces a
high increase in latency (perhaps of minutes) in the case of a
failure, since the results of the computation are never required to be
fresh.  The third regime is ``lazy checkpoint.'' This is used for the
real-time analytics subgraph which maintains complex state that must
be regularly checkpointed. In the event of a failure it is acceptable
to re-execute a few seconds' worth of work in this regime, so
checkpoints need not be taken every time state is updated. The final
regime is ``eager checkpoint.'' This is used for the database updates
which must be persisted as soon as they are recorded, since they must
be consistent with delivered results.  There exist fault tolerance
designs that fit several of these regimes, but no current system can
include them all in a single application as we desire.  The Falkirk
Wheel framework makes this flexible mixture of policies possible.

In common with prior work~\cite{alvisi:survey:acm02} we propose to
recover from a failure by restoring processes to
previously-checkpointed states, optionally replaying logged events
such as message deliveries that occurred after the checkpoints were
taken, then restarting execution. Many standard checkpointing and
logging techniques can be understood in terms of events tagged with
partially-ordered \emph{logical times}. After a failure the effect of
events at logical times in a chosen set is restored from saved state,
and events with times outside the set are re-executed.  This paper
makes two major contributions.  First, we show how different subgraphs
of a dataflow can make use of different logical time \emph{domains}.
This permits different styles of checkpointing, with different
performance tradeoffs, to coexist within a single fault-tolerant
application. We set down simple rules and a general algorithm for
choosing a consistent global state after a failure, taking into
account these different time domains.  Second, we introduce the
concept of \emph{selective rollback}. This means that a process that
has processed events at two different logical times $t_1$ and $t_2$
may be able to preserve the work for time $t_1$ after rollback but
undo and re-execute the work for $t_2$, independent of the order in
which the work was originally performed.  We show that selective
rollback allows new performance tradeoffs that are particularly
well-suited to high-throughput, low-latency systems such as
Naiad.

Our implementation targets the Naiad system, which previously had only
basic support for fault tolerance. Naiad adopts a single underlying
system mechanism and implements different computational models 
as libraries. Our design allows each library to adopt a checkpointing
policy tailored to its performance characteristics, while still
allowing the libraries to interact within a single application. Since
Naiad supports sophisticated streaming algorithms that may include
nested loops, it is a good testbed for general fault tolerance
mechanisms.  The ideas set out in this paper are applicable well
beyond Naiad, and their implementation in a system without cyclic
dataflow would be simpler.  For example, we believe that some of the
techniques we describe could be used, with modest effort, in the
context of the Spark Streaming system~\cite{zaharia:dstreams:sosp13}.

The next section sketches a number of popular fault tolerance policies
and explains selective rollback.  Section~\ref{sec:framework} sets out
the Falkirk Wheel design, and Section~\ref{sec:implementation}
describes its implementation in the Naiad system.
We finish with conclusions.

\section{Tracking events for rollback}
\label{sec:trackingevents}

\begin{figure*}[t]%
  \begin{center}%
    \subfloat[][Sequence numbers]{%
      \includegraphics[width=5cm]{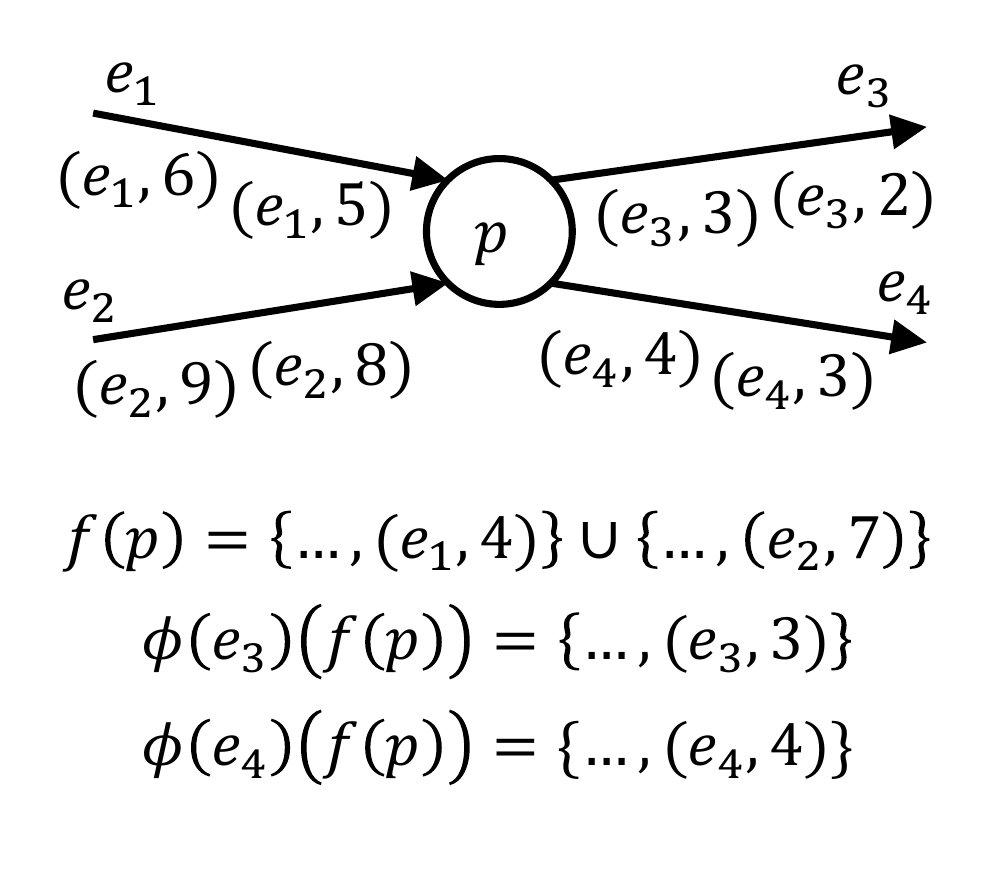}}%
    \subfloat[][Epochs]{%
      \includegraphics[width=5cm]{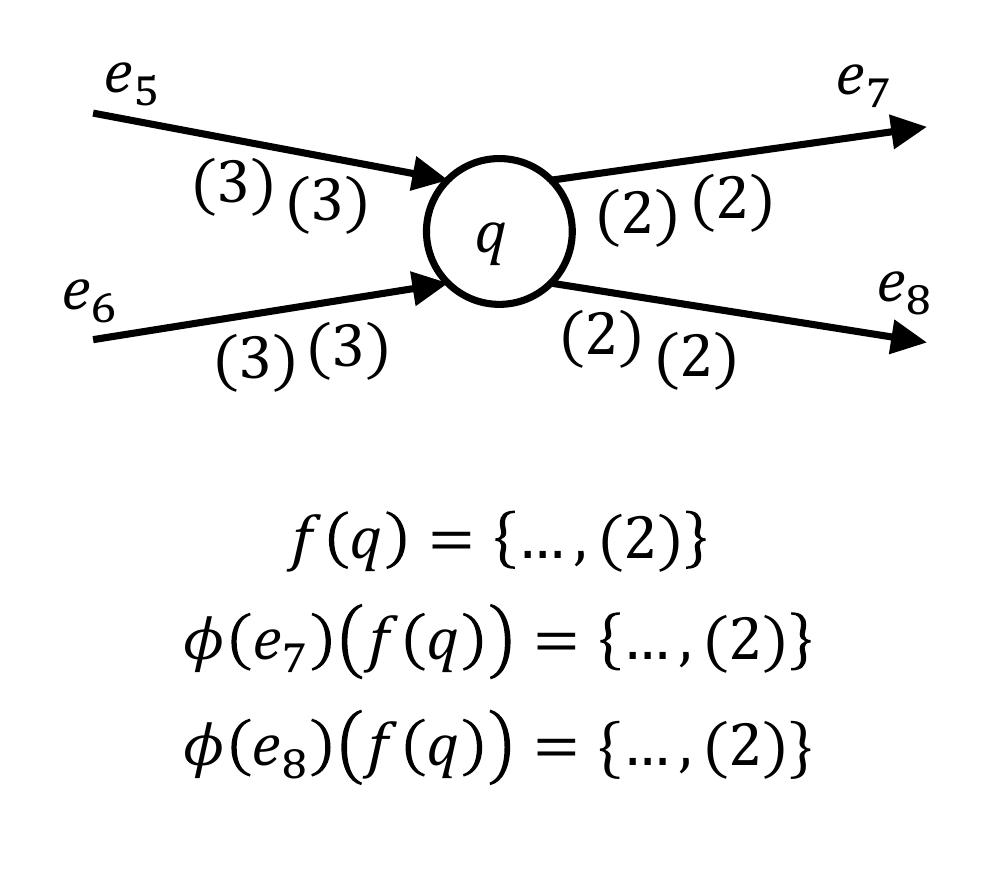}}%
    \subfloat[][Structured times: epochs entering a loop]{%
      \includegraphics[width=5cm]{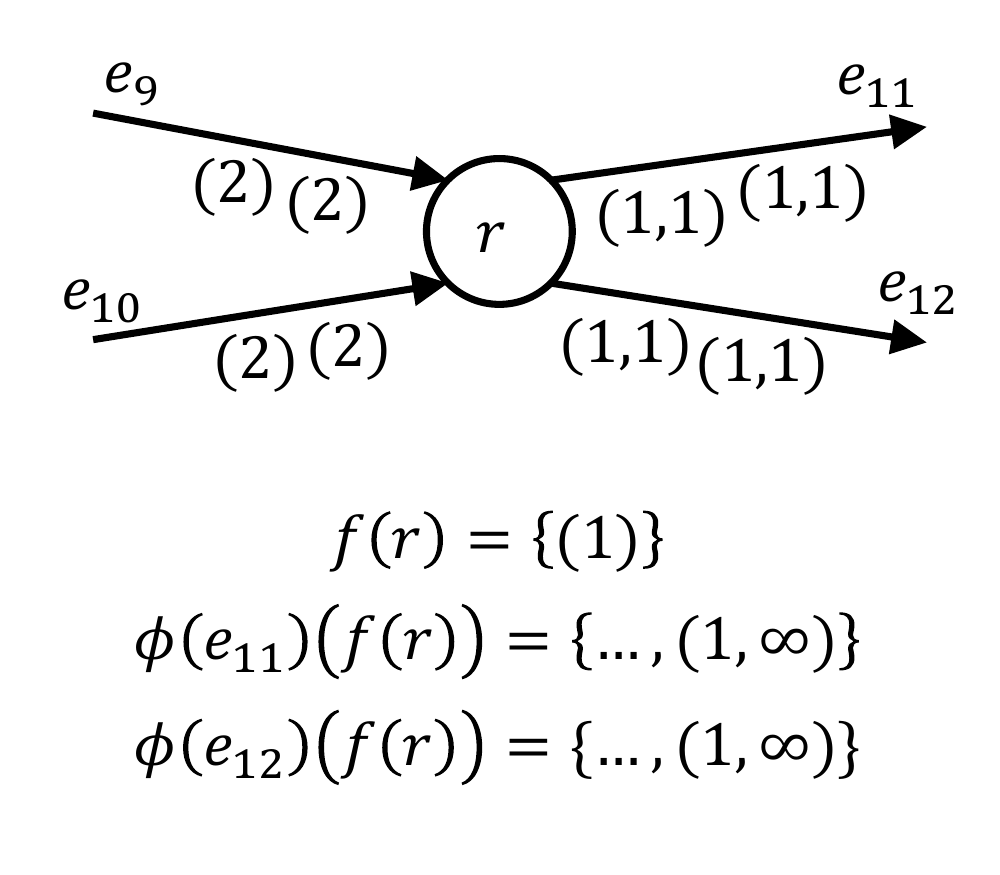}}%
  \end{center}%
  \caption{\textbf{Logical times for events.} Tuples
    $(\cdot)$ on input edges represent messages that have not yet been
    processed: the tuple shows the logical time of the message.
    Tuples on output edges represent sent messages.  In Scheme (a),
    the logical time of a message with sequence number $s$ on edge $e$
    is $(e,s)$. $p$ has processed the first 4 messages on edge $e_1$
    and the first 7 on $e_2$, and has sent 3 messages on $e_3$ and 4
    on $e_4$. Scheme (b) uses epoch numbers as logical times, so all
    messages in a given epoch have the same time. $q$ has processed
    all the events in the first two epochs, and has sent all
    corresponding messages for those epochs. Scheme (c) uses
    structured logical times, generalizing epochs. $r$ forwards
    incoming messages into a loop, which has a different time domain
    that includes an additional loop iteration counter. $r$ has
    processed all events in the first epoch and sent all the messages
    it will ever produce with epoch 1 and any iteration count. The
    \emph{frontier} $f(x)$ and \emph{edge projection} $\phi(e)(f(x))$
    at processor $x$ are discussed in Section~\ref{sec:framework}.}
  \label{fig:frontiers}
\end{figure*}
In this section we summarize a few rollback recovery schemes and
comment on the design and performance tradeoffs they embody. In our
discussion we refer to a processing node in a dataflow graph as a
\emph{processor}.  A physical CPU in a distributed system may host
multiple such processors.  Later, we will fit several of the schemes
into our common framework.  In order to do this it is helpful to think
of messages sent between processors as being tagged with
partially-ordered logical times; often these tags are implicit. Many
systems can inform a processor when it will not see any more messages
with a particular logical time $t$. We call this a \emph{notification}
at time $t$. An \emph{event} at time $t$ means the delivery of either
a message or a notification with that time. In the following we divide
logical times into two broad categories: sequence numbers; and
structured times, which include epochs.


\subsection{Sequence numbers}
\label{sec:sequencenumbers}

Sequence numbers on ordered channels are illustrated in
Figure~\ref{fig:frontiers}(a). There is no need for notifications when
using sequence numbers, since each message has a unique time.
Rollback schemes that we model using sequence numbers are often used
for systems where computation is not naturally structured using
epochs. Such schemes include the following:

\paragraph{Distributed Snapshots.}
Chandy and Lamport described a general algorithm for checkpointing an
arbitrary distributed system~\cite{chandy:distributedsnapshots:tcs85}.
Each process $p$ receives messages from other processes in the system
on a set of point-to-point channels $E(p)$. Periodically the system
performs a global checkpoint: it chooses, for each process $p$ and
channel $e \in E(p)$, a sequence number $s_e$, and records the state
$C_p$ of $p$ after all the messages up to $s_e$ have been delivered on
$e$ and no others.  The checkpoint also includes a sequence of
undelivered messages $M_e$ on each channel $e$. The design of the
algorithm ensures that the chosen $\{C_p\},\{M_e\}$ form a consistent
global system state.  Following a failure the system is restored to
the state at the most recently saved checkpoint. This scheme is
general, but has some practical drawbacks.  Each process must be able
to save a checkpoint at an arbitrary moment chosen by the system,
which introduces overhead that is side-stepped by some designs below.
Also in general all processes, even non-failed ones, must roll back to
a prior checkpoint following a failure.

\paragraph{Exactly-once streaming.}
Streaming systems including Storm~\cite{storm} and
Millwheel~\cite{akidau:millwheel:vldb13} support stateful processors
to which a message is guaranteed to be delivered exactly once,
corresponding to the ``eager checkpoint'' regime of
Figure~\ref{fig:scenario}. On receiving a message a processor persists
its updated state and any resulting outgoing messages before
acknowledging the processed message. As with the Chandy-Lamport
algorithm, the persisted state encodes the effect of processing all
messages up to the latest sequence number on each input, and no
others. If a processor fails it is restored to its most-recently
persisted state, which includes the effect of all acknowledged
messages. This scheme has several benefits: it allows processors to
choose locally when to checkpoint; it can guarantee high availability;
non-failed processors need never be interrupted; and processors may
join and leave the computation with low overhead since the system need
not keep track of the dataflow topology.  Drawbacks include a possible
throughput penalty because all mutations to state must be persisted,
and a possible latency penalty because sent messages must be
acknowledged by their recipient process before the next incoming
message can be acknowledged. The chain of dependent acknowledgements
that builds up as a message's effects propagate may also limit the
practical complexity of computations; for example iterative algorithms
may be problematic.\footnote{Millwheel addresses some latency concerns
  by partitioning the state at each processor by a key function and
  performing work for distinct keys in parallel. It can also notify a
  processor when a low-watermark has passed, based on wall-clock
  timestamps. These notifications are not the same as the logical-time
  notifications in this paper, and we can model them as messages
  delivered on a virtual edge.}

\paragraph{At-least once streaming.}
Both Storm and Millwheel also allow processors to be placed in a
relaxed fault tolerance mode, in which the system does not eagerly
checkpoint each state update before proceeding to the next. This gives
better performance, but must only be used for processors for which
message deliveries are idempotent, or where it is tolerable to end up
in a globally-inconsistent state. It is suitable for the ``ephemeral''
regime in our example.

\subsection{Epochs}
\label{sec:epochs}

Some systems associate each input message with a particular batch or
\emph{epoch}, and structure computation (often using dataflow) so that
all consequent messages and state updates can in turn be tagged with
an epoch. These epochs can be used as coarse-grain logical times for
events as illustrated in Figure~\ref{fig:frontiers}(b).

A number of recent acyclic batch dataflow
systems~\cite{dean:mapreduce:osdi04,isard:dryad:eurosys07,zaharia:spark:nsdi12}
share a fault tolerance model pioneered by the MapReduce
system~\cite{dean:mapreduce:osdi04}. Each processor reads all of its
inputs then implicitly receives a notification that the input is
complete, writes its outputs, empties its state, and quiesces.  We can
think of all inputs and messages as being in a single epoch $0$.  Each
system design specifies a subset of the edges in the dataflow and
persists the messages sent on those edges.  Following a failure the
system chooses to restore each failed processor either to the state
where it has processed no events, or where it has processed all
events, based on a global function of which sent messages have been
persisted.

This design has the appealing property that processors are always
restored to an empty state after failure: this means that the
(user-supplied) application logic in the processor need not include
any checkpointing code. On the other hand, any work in progress at the
time of a failure is lost and must be redone.  Non-failed processors
need only be interrupted if they have consumed messages from
processors that were restored to the empty state.  The model is well
suited to off-line data-parallel workloads, where throughput in the
absence of failures is paramount and delayed job completion is
tolerable in the event of failures.  A variation on the model, Spark
Streaming~\cite{zaharia:dstreams:sosp13}, allows each processor to
accept messages at an epoch $t+1$ after the messages at epoch $t$ have
been fully processed, matching the ``batch'' regime of our example.
Unlike traditional streaming systems it does not let processors retain
internal state between logical times.

\subsection{Selective rollback}
\label{sec:selectiverollback}

\begin{figure}[tbp]%
  \begin{center}%
    \leavevmode%
    \includegraphics[width=8cm]{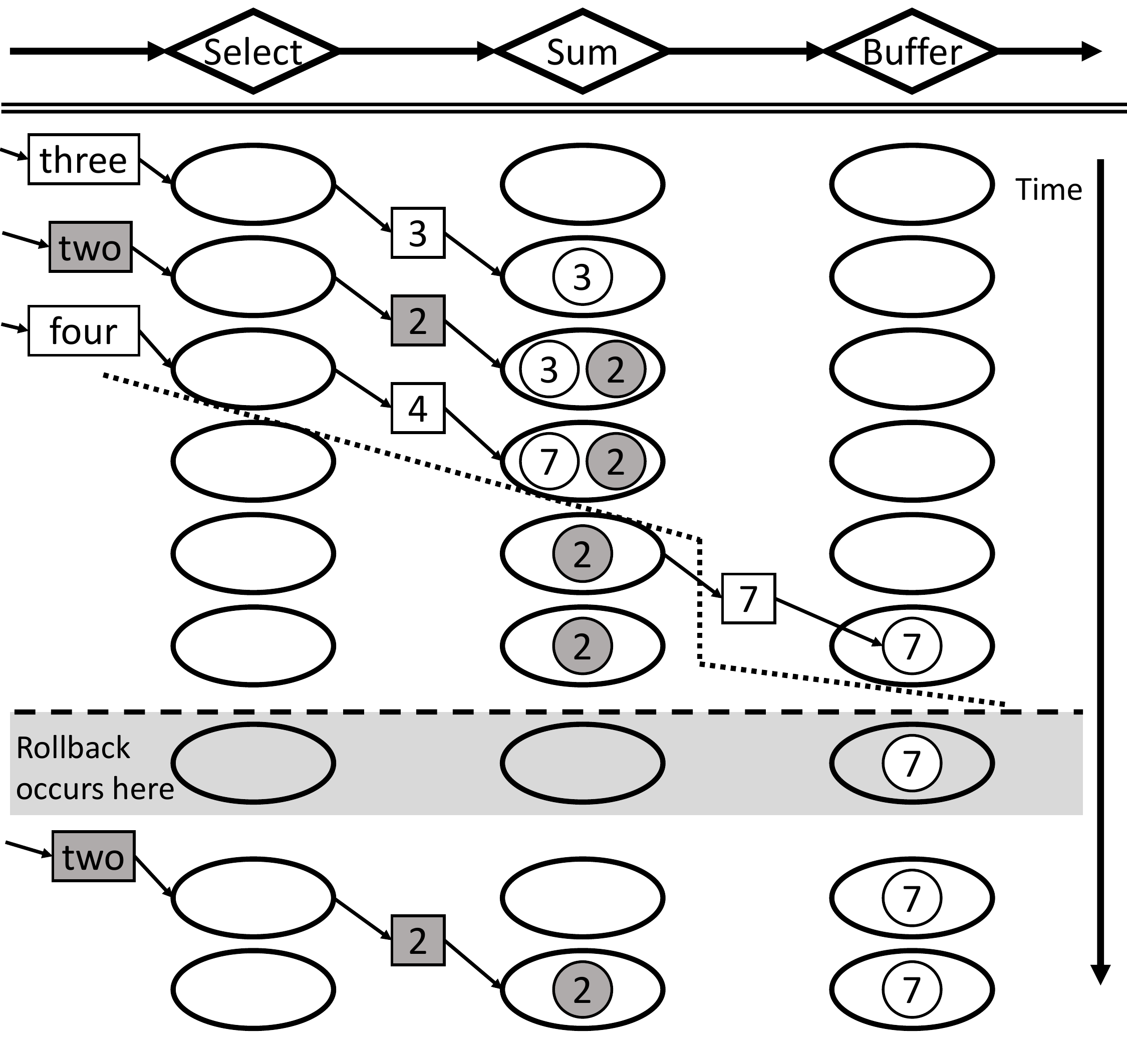}%
  \end{center}\vskip -3ex%
  \caption{\textbf{Selective rollback.} Rectangles show messages and
    ovals processor state. A white background indicates a message or
    state corresponding to logical time $A$; a grey background to time
    $B$. The dashed line shows the point at which a processor will not
    receive any more messages at time $A$; a notification is delivered
    to the Sum processor after this point, causing it to send a
    message and discard its state related to $A$. Processors roll back
    to a state where they have consumed all messages at $A$ and none
    at $B$.}
  \label{fig:selectiverollback}%
\end{figure}

The Naiad system~\cite{murray:naiad:sosp13} achieves state of the art
performance on the streaming iterative workload needed for the ``lazy
checkpointing'' regime of our example, so we consider its
fault-tolerance requirements. Naiad explictly assigns logical times to
events. Each time is a tuple indicating an input epoch along with loop
counters tracking progress through (possibly-nested) iteration as in
Figure~\ref{fig:frontiers}(c).  A processor can request that a
notification be delivered when a logical time is complete.
Figure~\ref{fig:selectiverollback} shows a fragment of a simple Naiad
dataflow graph made up of Select, Sum and Buffer processors.  Below it
is a timeline showing event deliveries and corresponding
updates to the processor state, colored according to logical times.
The Select processor translates a word into its numeric
representation, and is stateless. The Sum processor accumulates a
separate sum for each logical time. When notified that there will be
no more messages at a given time, Sum outputs the accumulated sum for
that time and then removes the sum from its local state. The Buffer
processor records all messages it has seen.

All the Naiad computational libraries developed so far, including
differential dataflow~\cite{mcsherry:diffdataflow:cidr13} which is the
most complex, either keep no state at a processor or partition its
state by logical time. Many Naiad processors, like the Sum in our
example, delete the state corresponding to a time once that time is
complete. It is thus desirable to allow a processor to wait until time
$t$ is complete before checkpointing the portion of local state that
corresponds to $t$. Often this means no checkpoint need be saved,
matching the software-engineering and performance characteristics of
the systems in Section~\ref{sec:epochs}.

Naiad applications often include loops implemented as distributed sets
of processors, and messages can flow around these loops with latencies
of a millisecond or less. Restricting Naiad to suspend delivery of a
message until all messages with earlier times had been processed would
force a processor to stall waiting for the global coordinator to
ensure that no ``earlier'' messages remained in the system,
introducing a severe performance penalty. Consequently, Naiad
processors may interleave the delivery of messages with different
logical times.

We introduce the idea of selective rollback in order to support
Naiad's twin performance requirements that processors must be able to
interleave the logical times of delivered messages, and also
checkpoint only state corresponding to completed times. In
Figure~\ref{fig:selectiverollback} each processor makes a selective
checkpoint after seeing the last time $A$ message. Rather than saving
its full current state, as is traditional, it saves the state it would
contain having seen all time $A$ messages and no time $B$ messages. In
general this checkpoint may not correspond to a state the processor
has previously been in.  The shaded rectangle shows a rollback during
which each processor is set to its checkpointed state.  Subsequently
an upstream processor is re-executed, causing the time $B$ message to
be re-sent, and eventually the state of the system returns to that
before the rollback. A scheme that did not support selective rollback
would be forced to prevent the interleaved delivery of messages at
different times, or to checkpoint non-empty state for the Sum
processor, either of which would introduce a substantial performance
penalty for Naiad.

\section{The Falkirk Wheel framework}
\label{sec:framework}

We now describe our general framework for rollback using logical
times. As previously mentioned, after a failure the system chooses a
set of logical times at each processor, which we call a
\emph{frontier}, and restores the processor to a state including the
effect of the previously-delivered events with times in that frontier.
We first discuss some restrictions on the use of logical times in our
framework, and show that the existing schemes described in
Sections~\ref{sec:sequencenumbers} and~\ref{sec:epochs} satisfy these
restrictions. We then discuss a general algorithm for choosing
frontiers that will result in rolling back to a globally consistent
state.

\subsection{From sets to frontiers}
\label{sec:downwardclosed}

Not all sets of logical times can be used as frontiers: a frontier
must be downward-closed. This means that if a time $t$ is in the
frontier, then so is every time $t' \le t$. For a set $T$ of times we
write ${\DClos}T = \{t' : t \in T \land t' \le t\}$ for the operation
that converts a set into the smallest frontier containing that set.
The schemes described in Sections~\ref{sec:sequencenumbers}
and~\ref{sec:epochs} already naturally adopt frontiers for rollback.
For epochs logical times are totally ordered, so the restriction
simply means that if we are rolling back to epoch $t$ we must also
include all previous epochs $t' < t$. For sequence numbers, recall
that a logical time is a pair $(e,s)$ where $e$ is an edge and $s$ is
the sequence number of a message on that edge. We define a partial
order on these times where $(e_1,s_1) \le (e_2,s_2)$ if and only if
$e_1 = e_2 \land s_1 \le s_2$. This means that times are only
comparable if they correspond to messages on the same edge, and within
an edge sequence numbers indicate the natural ordering.  For a
processor with incoming edges $e_1\ldots e_n$ we associate the state
in which the processor has consumed all messages up to $s_i$ on edge
$e_i$ with the set
\[
\begin{split}
  f^s_{e_1,\ldots,e_n}(s_1,\ldots,s_n) =&\\
  \{(e_1,1),\ldots,(e_1,&s_1)\} \cup \ldots \cup
  \{(e_n,1),\ldots,(e_n,s_n)\}.
\end{split}
\]
This set is a
frontier under the partial order above, and corresponds to the
messages whose effects are included in a checkpoint at that
state. Figure~\ref{fig:frontiers}(a) shows the frontier
$f(p)=f^s_{e_1,e_2}(4,7)$.

\subsection{Bridging time domains}
\label{sec:phifunctions}

The \emph{edge projection} functions $\phi(e)$ shown in
Figure~\ref{fig:frontiers} allow us reason about rollback in a system
containing processors with different logical time domains. For each
edge $e$ from processor $p$ to $q$, $\phi(e)(f)$ maps a frontier $f$
at $p$ to a frontier in the time domain of $q$. The function $\phi(e)$
must be consistent with the behavior of $p$: it is a conservative
estimate of the times that were ``fixed'' on $e$ given the events in
$f$ at $p$.  Specifically, $p$ is guaranteed not to have produced any
messages with times in $\phi(e)(f)$ as a result of processing an event
with a time outside $f$. Informally, this means it is ``safe'' to roll
$q$ back to $\phi(e)(f)$ as long as $p$ rolls back to a frontier at
least as large as $f$. We could always set $\phi(e)(f)=\emptyset$, but
instead would like to choose it as large as possible since larger
$\phi$ will allow us to preserve more work during rollbacks.

In rollback schemes that use sequence numbers $\phi(e)(f)$ is defined
naturally as illustrated in Figure~\ref{fig:frontiers}(a). Suppose
that when $p$ is in state $f^s_{e_1,\ldots,e_n}(s_1,\ldots,s_n)$
it has sent $s$ messages on outgoing edge $e$. Then
\[
\phi(e)(f^s_{e_1,\ldots,e_n}(s_1,\ldots,s_n)) =
\{(e,1),\ldots,(e,s)\}.
\]
(Conveniently, for our purposes we need not define $\phi(e)(f)$ for
any frontier that does not correspond to a state in the history of
$p$.)  Systems that use epochs typically adopt the restriction that
messages cannot be sent backwards in time. For these systems we can
set $\phi(e)(f)=f$ everywhere, meaning an event at epoch $t$ cannot
result in a message at any epoch $t'<t$.
\begin{figure*}[t]
  \begin{center}
    \leavevmode
    \includegraphics[width=15.5cm]{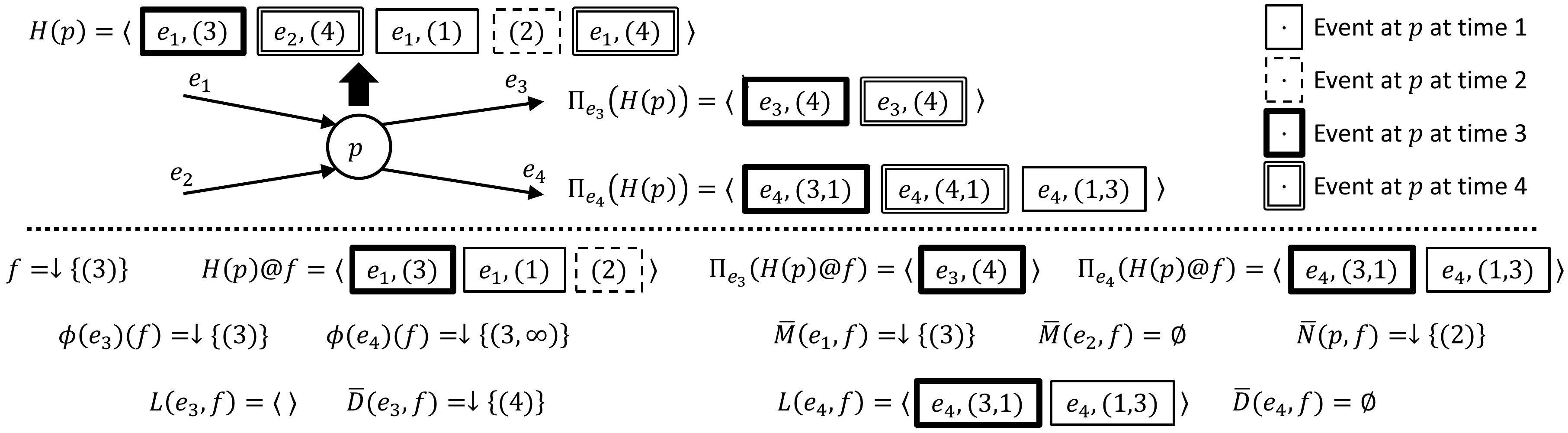}
  \end{center}\vskip -3ex
  \caption{\textbf{Processor $p$ filters its history on rollback.}
    $H(p)$ shows a sequence of events at $p$. Earlier events are to
    the left. There are three delivered messages, then a notification,
    then another message.  ${\msgon{e}}(H(p))$ shows the messages sent
    on edge $e$; so $p$ sent two messages with time $4$ on $e_3$.  The
    border around an event or sent message shows the time of the event
    \textbf{at $\mathbf{p}$}; so the first message on $e_3$ was sent
    at time $3$, and the second at $4$.  The state of $p$ after a
    rollback to the frontier $f=\{(1),(2),(3)\}$ is shown below the
    dotted line. The history and sent messages are filtered to retain
    only the events in $f$ at $p$.  ${\processedmess}(e_1,f)$,
    ${\processedmess}(e_2,f)$ and ${\processednot}(p,f)$ are the
    minimum frontiers containing the processed messages and
    notifications, respectively, in $p$'s filtered history. The
    processor logged all sent messages on $e_4$ and none on $e_3$.}
  \label{fig:filtering}
\end{figure*}

Figure~\ref{fig:frontiers}(c) shows an example of a processor that
receives messages tagged with epochs and forwards them in a new time
domain: sent messages have times $(t,c)$ where $t$ is the epoch of the
incoming message and $c$ a loop counter. In this case we can choose
$\phi(e)(f)$ to be $\{(t,c) : t \in f\}$, so $\phi$ ``translates''
between time domains.

Even in systems without loops, it may be useful to translate between
time domains. A processor $p$ may want to read from a computation
structured using epochs and forward its input to a processor that
takes eager checkpoints according to sequence numbers. In this case we
might require $p$ to forward all epoch $1$ data before sending any
epoch $2$ data, if necessary buffering epoch $2$ data until epoch $1$
is complete.  Suppose that in total $p$ receives $73$ messages in
epoch $1$, we could choose $\phi(e)(\{1\}) = \{1,\dots,73\}$. A
similar transformer could translate from sequence numbers to epochs,
for example to construct epochs from sets of messages received at a
processor within particular windows of wall-clock time.

\subsection{Message re-ordering}
\label{sec:reordering}

We must impose a restriction on the semantics of processors that will
be subject to selective rollback. This does not affect the schemes
described in Sections~\ref{sec:sequencenumbers} or~\ref{sec:epochs},
which never perform selective rollback. We require that such a
processor $p$ must be able to perform a limited re-ordering of
messages on its input edges.  Suppose $e$ is an input edge to $p$, and
contains a sequence of messages $\langle m_1,\ldots,m_k\rangle$ where
$m_1$ is at the head of the sequence, i.e.\ $m_1$ was sent before
$m_2$, and so on. Then $p$ is at liberty to choose to remove and
process from $e$ any message $m_i$ where ${\eime}(m_j) \not\le
{\eime}(m_i)~\forall j<i$. So if $m_5$ is in epoch $1$ and all of
$m_1\ldots m_4$ are in epochs $2$ or greater, $p$ can choose to
process $m_5$ next. It does not have to be the case the $p$ produces
the same output under all re-orderings, but all of the outputs have to
correspond to legal behaviors of the computation.  This restriction is
intuitively necessary if we want to legally be able to roll $p$ back
to a state in which it has processed all the epoch $1$ events and none
from later epochs, independent of the order that the messages appeared
on $e$. It is satisfied by all Naiad processors we are aware of.

\subsection{Checkpoints and processor history}
\label{sec:checkpoint}

When deciding what frontiers a processor can be rolled back to we need
to take into account exactly what information $p$ has persisted. For
example, processors can in general only roll back to a fixed set of
frontiers for which they took checkpoints. Also, some processors log
sent messages and others do not.

We start with notation. $H(p)$ is the \emph{history} at $p$ at the
time of the rollback, i.e.\ the sequence of events that it has
processed, and $H(p){\infilter}f$ is the subsequence of $H(p)$ keeping
only events with times in a frontier $f$. (For processors that don't
perform selective rollback, $H(p){\infilter}f$ is always a prefix of
$H(p)$.)  For $e \in {\Oute}(p)$, the output edges at $p$,
${\msgon{e}}(H(p))$ is the sequence of messages that $p$ sent on $e$
as a result of processing the events in $H(p)$, and
${\msgon{e}}(H(p){\infilter}f)$ is the sequence of messages that $p$
would have sent on $e$ if it had processed only the events in
$H(p){\infilter}f$. When $H(p){\infilter}f$ is not a prefix of $H(p)$,
${\msgon{e}}(H(p){\infilter}f)$ may not be a subsequence of
${\msgon{e}}(H(p))$, though it is for all the processors we have
studied. Figure~\ref{fig:filtering} shows an example history.

In general we don't have access to $H(p)$, ${\msgon{e}}(H(p))$,
$H(p){\infilter}f$, or ${\msgon{e}}(H(p){\infilter}f)$.  Instead, we
assume that there is some sequence of frontiers $F^*(p) =
\{f_1,\ldots,f_n\}$, where $f_i \subset f_{i+1}$, that are available
for $p$ to roll back to because it has persisted appropriate
information about them, summarized in Table~\ref{tab:persistedstate}.
For a processor that has not failed, $F^*(p)$ may contain the special
frontier $\top$ that includes all event times.
\begin{table}[t]
\centering
{
\begin{tabular}{ll}
$F^*(p)$ & Set of available frontiers\\[1ex]
\multicolumn{2}{l}{\textit{For each $f \in F^*(p)$}}\\
$S(p,f)$ & Internal state at $f$\\
${\processednot}(p,f)$ & Processed notification frontier at $f$\\[1ex]
\multicolumn{2}{l}{\textit{For each $f \in F^*(p)$, $d \in
    {\Ine}(p)$}}\\
${\processedmess}(d,f)$ & Processed message frontier from $d$ at $f$\\[1ex]
\multicolumn{2}{l}{\textit{For each $f \in F^*(p)$, $e \in
    {\Oute}(p)$}}\\
$\phi(e)(f)$ & Edge projection on $e$ at $f$\\
$L(e,f)$ & Messages logged on $e$ at $f$\\
${\discardedmess}(e,f)$ & Discarded message frontier on $e$ at $f$
\end{tabular}
}
\caption{\textbf{State that must be available to processor $p$ on
    rollback.} Most processors can approximate some values and do not
  need to explicitly persist all of them.}
\label{tab:persistedstate}
\end{table}

For each $f \in F^*(p)$ we assume $p$ has persisted enough
information to recover $\phi(e)(f)$ for each $e \in {\Oute}(p)$ and to
be able to restore its internal state to $S(p)(f)$, which reflects the
effects of all events in $H(p){\infilter}f$.  Depending on $p$'s
policy it may have logged some, all, or none of its sent messages. We
write $L(e,f)$ for the subsequence of ${\msgon{e}}(H(p){\infilter}f)$
that have been logged and
$D(e,f)={\msgon{e}}(H(p){\infilter}f)~\backslash~L(e,f)$ for those
that were discarded.  Let $M(d,f)$ be the sequence of messages on $d
\in {\Ine}(p)$, the input edges to $p$, that were processed by $p$ in
$H(p){\infilter}f$, and $N(p,f)$ the sequence of notifications
processed by $p$ in $H(p){\infilter}f$.  For each $f \in F^*(p)$ we
assume that $p$ has stored a conservative estimate of
${\processedmess}(d,f)$, ${\processednot}(p,f)$, and
${\discardedmess}(e,f)$, respectively the smallest frontier containing
its delivered messages, notifications, and discarded messages:
\[
\begin{split}
{\processedmess}(d,f) &=
  {\DClos} \{ t : (d,m) \in M(d,f) \land t = {\eime}(m) \} \\
  {\processednot}(p,f) &=
  {\DClos} \{ t : t \in N(p,f) \}\\
  {\discardedmess}(e,f) &=
  {\DClos}\{ t : m \in D(e,f) \land t = {\eime}(m) \}.
\end{split}
\]
Note that ${\eime}(m)$ for $m \in D(e,f)$, and thus also
${\discardedmess}(p,f)$, is in the domain of the process that will
\emph{receive} the message, not $p$'s time domain.

In many cases, $p$ need not explicitly store all the state in
Table~\ref{tab:persistedstate}. For most schemes that use structured
times, including epochs, $\phi(e)(f)$ is independent of $p$'s history.
It is always safe to overestimate
${\processedmess}(d,f)={\processednot}(p,f)=f$.  If the processor logs
all messages, ${\discardedmess}(e,f) = \emptyset$. For most processors
that discard all messages it is safe to use the approximation
${\discardedmess}(e,f) = \phi(e)(f)$, though processors that send
``into the future,'' like some differential dataflow
processors~\cite{mcsherry:diffdataflow:cidr13}, must explicitly keep
track of which times they have discarded messages for.  Finally, in
the common case (as in Section~\ref{sec:epochs}) that $p$ is an
epoch-based processor that keeps no state between epochs, sends all
messages with the epoch of the event that caused the message, and
doesn't log any messages, it need not persist anything. Such
processors can adopt
\[
\begin{split}
  S(p,f)&=\emptyset~~~~~~~~~~~L(e,f)=\langle\rangle\\
  \phi(e)(f)={\processedmess}(&d,f)=
  {\processednot}(p,f)={\discardedmess}(e,f)=f
\end{split}
\]
and need not even save $F^*(p)$ since they can restore to any
requested frontier.

\subsection{Consistent frontiers for rollback}
\label{sec:consistentrollback}

In the event of one or more failures, the system must choose a
frontier $f(p)$ at each processor $p$ such that the system as a whole
rolls back to a consistent global state. We list a set of constraints
that, if satisfied, ensure a consistent rollback. We have published a
theoretical paper that proves
the correctness of the constraints. We show via a refinement mapping
that a system which obeys the Falkirk Wheel rollback constraints on
failure implements (has external effects indistinguishable from) a
higher-level system without failures.

The first constraint says that a processor $p$ may not restore to a
frontier $f$ if there is any message $m$ awaiting delivery on an edge
$e \in {\Ine}(p)$ with ${\eime}(m) \in f$. This restriction can be
satisfied by saving a checkpoint for frontier $f$ only after all the
times in $f$ are complete at $p$. This behavior is already adopted by
the systems described in Sections~\ref{sec:sequencenumbers}
and~\ref{sec:epochs} and is easy to enforce for systems such as Naiad
that support notification.

The next constraint deals with discarded messages:
\[
\forall e \in {\Oute}(p),~~
{\discardedmess}(e,f(p)) \subseteq f({\edst}(e))
\]
where ${\edst}(e)$ is the processor that $p$ sends to on $e$.
Informally, this says that a processor downstream of $p$ cannot roll
back so far that it would need to re-receive any messages that $p$ has
discarded.

The third constraint deals with delivered messages:
\[
\forall d \in {\Ine}(p),~~
{\processedmess}(d,f) \subseteq \phi(d)(f({\esrc}(d)))
\]
where ${\esrc}(d)$ is the processor that sends to $p$ on $d$. This
says that a processor must roll back far enough that any delivered
messages are within the frontier ``fixed'' by the upstream processor's
rollback, in the sense described in Section~\ref{sec:phifunctions}.

\begin{figure}[tbp]
  \begin{center}
    \leavevmode
    \includegraphics[width=8cm]{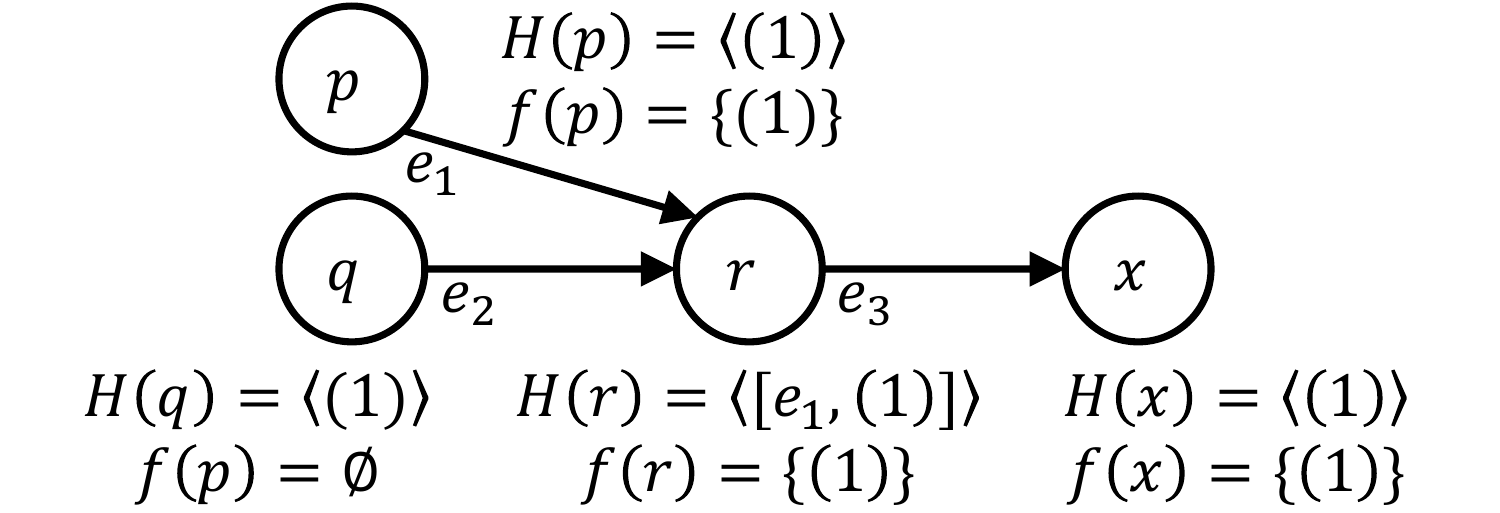}
  \end{center}\vskip -3ex
  \caption{\textbf{Without notification frontiers rollback can lead to
      inconsistent state.} See Section~\ref{sec:consistentrollback}
    for details.}
  \label{fig:notificationfrontier}
\end{figure}
The final constraints deal with notifications and are motivated by the
example in Figure~\ref{fig:notificationfrontier}, in which
$\phi(e)(f)=f$ for all $e$. Processors $p$ and $q$ have each received
a notification for time $1$, in response to which $p$ sent a message
at time $1$ on $e_1$ and $q$ did nothing. The message arrived at $r$,
which sent nothing in response, at which point $x$ received a
notification for time $1$ indicating that it will not receive any more
time $1$ messages. According to the preceding constraints the system
could roll back to the frontiers shown in the Figure; in particular
$f(p)$ can be set to $\emptyset$ since
${\processedmess}(e_2,\{(1)\})=\emptyset$. Suppose, after rollback,
$q$ behaves differently on receiving the notification and sends a
message at time $1$ on $e_2$, which $r$ forwards on $e_3$. Then $x$
will receive a new message at time $1$ even though it has rolled back
to a history in which it received a notification promising this will
never happen. The problem cannot be fixed by simply adding a new
pairwise constraint between $r$ and $x$: in the example they already
roll back to the same frontier. Instead we introduce an auxiliary
variable at each $p$, the \emph{notification frontier} $f_n(p)$, and
add additional constraints:
\[
\begin{split}
  f_n(p) &\subseteq f(p)\\
  {\processednot}(p, f) &\subseteq f_n(p)\\
  \forall d \in {\Ine}(p),~~
  f_n(p) &\subseteq \phi(d)(f_n({\esrc}(d))).
\end{split}
\]
The notification frontiers are not used in the rollback; they simply
act to constrain $f(p)$ to ensure consistency. Notification frontiers
can be ``omitted'' by setting ${\processednot}(p,f)=f_n(p)=\emptyset$
everywhere in systems without notifications.

\subsection{Choosing consistent frontiers}
\label{sec:choosingfrontiers}

Figure~\ref{fig:rollbackalgorithm} shows an algorithm to find a
frontier at each processor that will satisfy the constraints. As long
as $\emptyset \in F^*(p)~\forall~p$, meaning every processor can roll
back to its initial state, it is always possible to choose values for
$f'$ and $f_n'$ while executing the fixed point. In this case the
algorithm will always converge since neither $f$ nor $f_n$ ever
increases, and $f(p)=f_n(p)=\emptyset~~\forall~p$ satisfies all
constraints.

The choice of $f'_n(p)$ indicates a maximum over a subset of all
frontiers.  If frontiers are not totally ordered, any maximal element
can be chosen. In all practical systems we have considered either
frontiers are totally ordered, notifications are not supported, or
${\processednot}(p,f)=f(p)$ everywhere (so $f_n(p)=f(p)$). In such
systems the algorithm will at every $p$ return the maximal
globally-consistent frontier (and the term ${\processednot}(p,f'(p))
\subseteq g_n$ is unnecessary since it will always be satisfied). For
these systems, adding choices of $f$ to $F^*(p)$ at any $p$ will never
cause $f(p')$ to get smaller for any $p'$---a valid set of frontiers
remains valid as more checkpoints are saved.

\begin{figure*}[t]%
  \begin{center}%
    \leavevmode%
    \includegraphics[width=16cm]{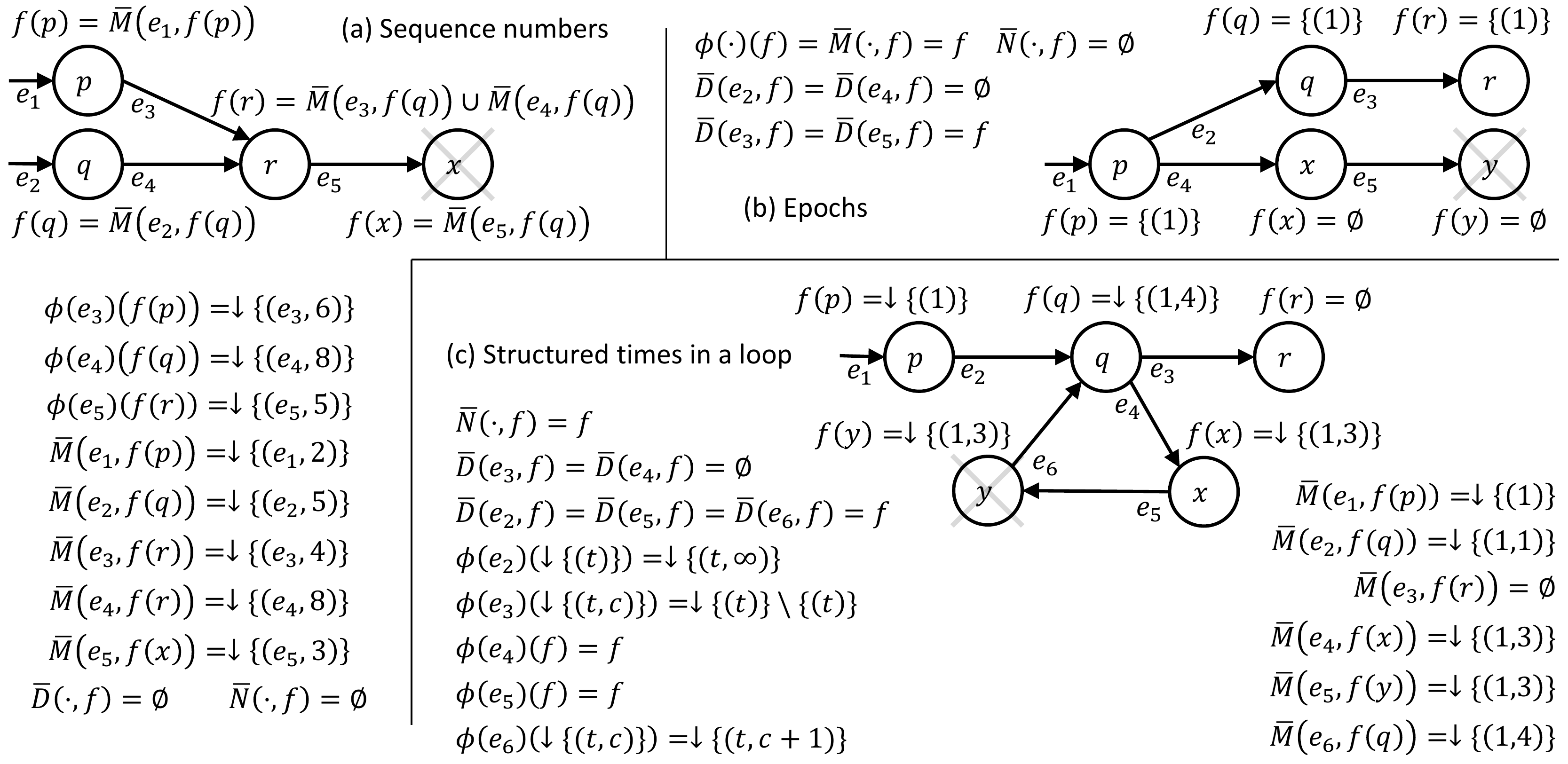}%
  \end{center}\vskip -3ex%
  \caption{\textbf{Some examples of rollback.} Panel (a) shows a
    system based on sequence numbers. Processor $x$ has failed. All
    processors log all outputs (${\discardedmess}(\cdot,f)=\emptyset$)
    and there are no notifications. All processors roll back to a
    state where they have sent at least as many messages as their
    upstream processors have consumed. Panel (b) shows a system based
    on epochs, similar to Spark~\cite{zaharia:spark:nsdi12}, where $y$
    has failed. Processor $p$ acts like a Spark Resilient Distributed
    Dataset (RDD) and has logged all its outputs; no other processors
    have saved any state. Both $x$ and $y$ must roll back to their
    initial state, while $p$, $q$ and $r$ do not need to roll back.
    Panel (c) shows a system like Naiad with a loop, where $y$ has
    failed.  Processor $q$ logs its sent messages, but no other
    processors do.  Processor $p$ sends messages into the loop along
    $e_2$ and $q$ sends them out of the loop along $e_3$.  Processor
    $q$ increments the loop counter coordinate of the time of each
    message it receives on $e_5$, and then forwards it on $e_6$. As a
    result, $q$ can roll back to ${\DClos}\{(1,4)\}$ even though $y$
    rolls back to ${\DClos}\{(1,3)\}$. Thus $q$ re-sends its logged
    messages at time $(1,4)$ on $e_4$, ``restarting'' the
    processing in the loop.}%
  \label{fig:examplerollback}%
\end{figure*}

After frontier $f(p)$ is chosen for rollback at $p$, its state is reset
as follows:
\[
\begin{split}
  {F^*}'(p) &= \{ f' : f' \in F^*(p) \land f' \subseteq f(p) \}\\
  H'(p) &= H(p){\infilter}f(p)\\
  S'(p) &= S(p,f(p))\\
  Q'(e) &= L(p,f(p)){\outfilter}f({\edst}(e))~~\forall e \in {\Oute}(p)
\end{split}
\]
where $Q'(e)$ is a sequence of messages to send on $e$ and
$L(p,f(p)){\outfilter}f({\edst}(e))$ is the messages in $L(p,f(p))$
whose times are not contained in
$f({\edst}(e))$. Figure~\ref{fig:examplerollback} shows some examples
of dataflow graphs with different characteristics, and the frontiers
that are chosen for rollback.
\begin{figure}[t]
\paragraph{Initially:} $\forall p,~~f(p)=f_n(p)=\max\{f \in F^*(p)\}$.
\vspace{2ex}
\paragraph{Continue until fixed point:}
\begin{equation*}
\begin{split}&\\[-4ex]
  f'&(p)=\max\{g \in F^*(p) \textrm{ such that } g \subseteq f(p)\\
  &{}\land \forall e \in {\Oute}(p),~~
  {\discardedmess}(e,g) \subseteq f({\edst}(e))\\
  &\begin{split}
    {}\land \forall d \in {\Ine}(p),~~
    {\processedmess}(d,g) &\subseteq \phi(d)(f({\esrc}(d)))\\
    {}\land {\processednot}(p,g) &\subseteq \phi(d)(f_n({\esrc}(d)))\}
  \end{split}\\[1ex]
  f_n'&(p)=\max\{g_n \textrm{ such that } g_n \subseteq f'(p) \cap f_n(p)\\
  &~~~~~~~~~~~\begin{split}
    &{}\land {\processednot}(p,f'(p)) \subseteq g_n\\
    &{}\land \forall d \in {\Ine}(p),~~
    g_n \subseteq \phi(d)(f_n({\esrc}(d)))\}
  \end{split}
\end{split}
\end{equation*}
  \caption{\textbf{Algorithm to choose consistent frontiers for rollback.}}%
  \label{fig:rollbackalgorithm}%
\end{figure}

\section{Fault tolerance in Naiad}
\label{sec:implementation}

In order to evaluate its performance and ease of use, we have added
prototype support for Falkirk Wheel fault tolerance to
Naiad~\cite{murray:naiad:sosp13}. Naiad is structured as a low-level
system layer, a set of commonly-used framework libraries, and a few
application-specific processors.  The Lindi framework is a library of
processors that keep no state between logical times, with similar
functionality to Spark~\cite{zaharia:spark:nsdi12} plus native support
for iteration.  Differential
Dataflow~\cite{mcsherry:diffdataflow:cidr13} is a general-purpose
library for incremental iterative computation, in which processors
generally keep state to allow them to respond quickly to updates. As
we explain in the following, we have added appropriate checkpointing
and logging to all the Lindi and Differential Dataflow processors, as
well as hooks to make it easy to add fault tolerance to custom
processors.

\subsection{Logging and checkpointing support}
\label{sec:processorlogging}

For simplicity, for checkpointing purposes we impose the lexicographic
(total) ordering on all Naiad logical times at a given
processor.
Since logical times at a processor are totally ordered a frontier can
be summarized by a single largest element, and frontiers are also
totally ordered.

Naiad already requires that messages are serializable in order to
support distributed operation.  Any processor, with no additional
programming effort, can request that the system log all of its
delivered messages and notifications; i.e., its full history $H(p)$ in
the notation of Section~\ref{sec:framework}.  This gives any
deterministic processor without external side-effects full fault
tolerance with no software-engineering effort: it can be automatically
rolled back to any frontier by replaying the filtered history and
forwarding any resulting messages that are needed by downstream
processors after their rollback. This is a good fallback option, but
the history grows without bound so it is not suitable for long-running
streaming applications.

The system can automatically keep track of ${\processednot}$,
${\processedmess}$ and ${\discardedmess}$ for any processor. A
processor can elect to log some or all sent messages, again with no
additional programming effort. A processor can also declare that it
keeps no state between logical times, and we call such a processor
``stateless'' even though it may accumulate state within a time.
Alternatively it can elect to receive checkpoint callbacks. If such a
processor requests a notification for time $t$ then it may selectively
checkpoint its state up to $t$ after the notification has been
processed. Stateful processors are also periodically (lazily) informed
when new times become complete, and can choose to selectively
checkpoint based on local policy.

We identify all Lindi processors as stateless, and by default suppress
logging of sent messages meaning that the processors incur no fault
tolerance overhead. A particular instance of a processor may be told
by an application developer to log its sent messages, in which case it
behaves like a Spark RDD and acts like a ``firewall'' preventing
upstream processors from rolling back in the event of a downstream
failure.

We have added selective incremental checkpointing to all Differential
Dataflow processors that keep state. Since the state is internally
stored differentiated by logical time, this was straightforward.

\subsection{Garbage collection}
\label{sec:garbagecollection}

A fault tolerance design that targets practical streaming systems must
address the issue of garbage-collecting persisted state, since it will
otherwise grow indefinitely. Let
\begin{multline*}
{\CMetadata}(p,f) = \bigl\{ f, {\processednot}(p,f),\bigr.\\
\bigl.  \{{\processedmess}(d,f): d \in {\Ine}(p)\},
  \{ {\discardedmess}(e,f): e \in {\Oute}(p) \} \bigr\}
\end{multline*}
be the metadata about the checkpoint needed
for the rollback algorithm.

Each time a processor $p$ receives an acknowledgement from storage
that ${\CMetadata}(p,f)$, $S(p,f)$ and $L(p,f)$ have all been
persisted for some $f$, it sends ${\CMetadata}(p,f)$ to a monitoring
service. This service keeps track of $F^*(p)$ for all processors in
the system. It starts with $F^*(p)=\emptyset$ and updates it every
time it receives new metadata.  The monitor runs an incremental
implementation of the fixed point algorithm of
Figure~\ref{fig:rollbackalgorithm} in a local Naiad runtime
independent of the main application. When an update arrives the
algorithm determines the new maximum rollback frontier at every
processor given the persisted checkpoints. We assume that storage is
reliable, so this rollback frontier is a low-watermark: the processor
will never need to roll back beyond it in any failure scenario. Every
time the low-watermark frontier at $p$ increases to $f$ the monitoring
service informs $p$, which is at liberty to garbage-collect
${\CMetadata}(p,f')$ and $S(p,f)$ for any $f' \subset f$. Processors
$q$ that send to $p$ are also notified, and can discard any messages
in $L(e,\cdot)$ with times in $f$ for $e \in {\Ine}(p)$. Since the
monitoring service is deterministic, monotonic, and used only for
garbage collection, it could easily be replicated though our prototype
does not do this.

\subsection{Inputs and outputs}
\label{sec:inputoutput}

The fault-tolerance properties of a streaming system can only be
considered in the context of its streaming inputs and outputs. We
assume that the services producing and consuming streams support fault
tolerance via acknowledgement and retry. For an input, this means that
the service will keep a batch of data available, and re-send if
requested, until the batch has been acknowledged. For an output this
means that we must be willing to re-send a batch of data multiple
times until it is acknowledged by the recipient. These assumptions are
compatible with services such as Kafka~\cite{kafka} and Azure Event
Hubs~\cite{eventhubs}.

Input and output acknowledgements can be handled by our existing
garbage-collection mechanism. Processors that read external inputs are
marked as stateless. Once such a processor is informed by the monitor
that it will never need to roll back beyond a frontier $f$ it can
acknowledge all inputs ingested at times in $f$.  A processor that
sends external outputs is marked stateful but saves no checkpoints;
instead it tells that monitor that $f$ has been persisted once the
external service has acknowledged all records sent at times in $f$, at
which point the rest of the system may discard state that would be
needed to regenerate those output records.  We can use this mechanism
to construct a stateless pipeline in which input records are only
acknowledged once outputs have been consumed; or by adding persistent
state in the pipeline we can decouple input receipt from output
acknowledgement.

\subsection{Recovery from failure}
\label{sec:naiadrollback}

A processor $p$ typically discovers the failure of another processor
$q$ by the failure of a network connection to a remote computer. When
this happens $p$ continues to work, buffering output to $q$ in case
the connection is reestablished. When $q$'s failure is confirmed by a
failure detector, the system pauses all processors and uses the
monitoring service to determine appropriate rollback frontiers. All
non-failed processors $p$ have $\top$ temporarily added to $F^*(p)$,
and the incremental algorithm computes the maximal frontiers needed
for rollback given the failed processors. A non-failed processor with
a frontier earlier than $\top$ can typically roll back by discarding
in-memory state rather than restoring from stable storage.  Any needed
logged messages $Q'(e)$ are placed in appropriate output queues, and
the processors are restarted. With some additional work Naiad could be
modified to allow pipelines of non-failed processors to continue
without pausing.

\section{Conclusions}
\label{sec:conclusions}

We present a new framework for rollback recovery, suitable for high
throughput streaming systems. We show a general mechanism to determine
a globally consistent state given a collection of local checkpoints
and logs organized in terms of logical times, and information about
the local behavior of processors that constrains what logical times
may be assigned to messages sent in response to events. The generality
of the mechanism makes it possible for processors to use flexible
local policies to decide when to take checkpoints, and as a result get
substantial performance and software engineering benefits.



\bibliographystyle{abbrv}
\bibliography{defs,naiad}
\end{document}